\documentclass[usenatbib,usegraphicx]{mn2e}
\usepackage{amsfonts}
\usepackage{amsmath}
\usepackage{amssymb}
\usepackage{comment}
\usepackage{color}

\title[The birth of spheroids in the early Universe]
    {Newborn spheroids at high redshift: when and how did the dominant, old stars in today's massive galaxies
    form?}

\author[Sugata Kaviraj et al.]
{S. Kaviraj$^{1,2,3}$, S. Cohen$^{4}$, R. S. Ellis$^{3}$, S.
Peirani$^{5}$, R. A. Windhorst$^{4}$,
\newauthor  R. W. O'Connell$^{6}$, J. Silk$^{5,2}$, B. C. Whitmore$^{7}$, N. P.
Hathi$^{8}$, R. E. Ryan Jr$^{7}$,
\newauthor M. A. Dopita$^{9,10}$, J. A. Frogel$^{10,11}$ and A. Dekel$^{12}$\\\\
$^{1}$Blackett Laboratory, Imperial College London, London SW7 2AZ, UK\\
$^{2}$Department of Physics, University of Oxford, Keble Road,
Oxford, OX1 3RH, UK\\
$^{3}$California Institute of Technology, 105-24 Astronomy,
Pasadena, CA 91125, USA\\
$^{4}$School of Earth and Space Exploration, Arizona State
University, Tempe, AZ 85287-1404, USA\\$^{5}$Institut
d'Astrophysique de Paris, 98 bis boulevard Arago,
75014 Paris\\
$^{6}$Department of Astronomy, University of Virginia,
Charlottesville, VA 22904-4325, USA\\
$^{7}$Space Telescope Science Institute, Baltimore, MD 21218,
USA\\
$^{8}$Carnegie Observatories, 813 Santa Babara Street, Pasadena,
California, 91101, USA\\
$^{9}$Research School of Physics and Astronomy, The Australian
National University, ACT 2611, Australia\\
$^{10}$King Abdulaziz University, Astronomy Department, Faculty of
Science, Jeddah, Saudi Arabia\\
$^{11}$Galaxies Unlimited, 1 Tremblant Court, Lutherville, MD 2109
USA\\
$^{12}$Racah Institute of Physics, The Hebrew University,
Jerusalem 91904, Israel\\\vspace{-0.15in}}

\begin{document}

\maketitle

\def \aj {AJ}
\def \mnras {MNRAS}
\def \pasp {PASP}
\def \apj {ApJ}
\def \apjs {ApJS}
\def \apjl {ApJL}
\def \aap {A\&A}
\def \nat {Nature}
\def \araa {ARAA}
\def \iaucirc {IAUC}
\def \aaps {A\&A Suppl.}
\def \qjras {QJRAS}
\def \na {New Astronomy}
\def\lesssim{\mathrel{\hbox{\rlap{\hbox{\lower4pt\hbox{$\sim$}}}\hbox{$<$}}}}
\def\gtrsim{\mathrel{\hbox{\rlap{\hbox{\lower4pt\hbox{$\sim$}}}\hbox{$>$}}}}


\begin{abstract}
We study $\sim$330 massive (M$_*>10^{9.5}$ M$_{\odot}$), newborn
spheroidal galaxies (SGs) around the epoch of peak star formation
($1<z<3$), to explore the high-redshift origin of SGs and gain
{\color{black}insight} into when and how the old stellar
populations that dominate today's Universe formed.
{\color{black}The sample is drawn from the HST/WFC3 Early-Release
Science programme, which provides deep 10-filter (0.2 - 1.7
$\mu$m) HST imaging over a third of the GOODS-South field}. We
find that {\color{black}the star formation episodes that built
{\color{black}our} SGs} likely peaked in the redshift range
$2<z<5$ (with a median of $z\sim3$) and have decay timescales
shorter than {\color{black}$\sim$1.5 Gyr}. Starburst timescales
and ages show no trend with stellar mass in the range $10^{9.5}$
M$_{\odot}<$ M$_*<10^{10.5}$ M$_{\odot}$. However, the timescales
show increased scatter towards lower values ($<0.3$ Gyr) for
M$_*>10^{10.5}$ M$_{\odot}${\color{black},} and an age trend
becomes evident in this mass regime: {\color{black}SGs with
M$_*>10^{11.5}$ M$_{\odot}$ are $\sim$2 Gyrs older than their
counterparts with M$_*<10^{10.5}$ M$_{\odot}$}. Nevertheless, a
\emph{smooth} downsizing trend with galaxy mass is not observed,
and the large scatter in starburst ages indicates that SGs are not
a particularly coeval population. Around half of the \emph{blue}
SGs appear not to drive their star formation via major mergers,
and those that have experienced a recent major merger, show only
modest enhancements ($\sim$40\%) in their specific star formation
rates. Our empirical study indicates that processes other than
major mergers (e.g. violent disk instability driven by cold
streams and/or minor mergers) likely play a dominant role in
building SGs, and creating {\color{black} a significant fraction
of the} old stellar populations that dominate today's Universe.
\end{abstract}




\section{Introduction}
Massive galaxies are central to our understanding of the visible
Universe. Locally, the massive galaxy census is dominated by
systems with spheroidal morphology. Hosting more than $\sim$50\%
of the stellar mass at the present day
\citep[e.g.][]{Bernardi2003}, these massive spheroidal galaxies
(SGs) are key laboratories for studying the evolution of galaxies
over cosmic time. Their red optical colours
\citep[e.g.][]{BLE92,Ellis97,Stanford98,Gladders98,Bernardi2003,Bell2004,Kaviraj2005,Faber2007},
obedience {\color{black}of} a `fundamental plane' with little
intrinsic scatter \citep[e.g.][]{Jorg1996,Franx1995,Saglia1997,
Forbes1998,Peebles2002} and chemical evidence for short
star-formation timescales \citep[e.g.][]{Trager2000a,Thomas2005},
indicate that the bulk of their stellar mass formed rapidly at
high redshift ($z>1$). Nevertheless, the {\color{black}young
(luminosity-weighted) ages} observed in some SGs are a signature
of recent {\color{black}($<5$ Gyr)} star formation
\citep[e.g.][]{Ellis2001,Menanteau2001a,Nelan2005,Schiavon2007,Trager2008,Graves2009},
which can be accurately quantified using a sensitive probe of
young stars, such as the rest-frame {\color{black}ultraviolet (UV;
1200-3000 \AA)}. Recent UV studies have shown that, while old
stars do dominate today's SGs, a significant minority ($\sim$20\%)
of the stellar mass in these galaxies is formed at $z<1$ (Kaviraj
et al. 2007a; 2008, see also Ferreras \& Silk 2000a; Yi et al.
2005; Salim \& Rich 2010; de la Rosa et al. 2011; Rutkowski et al.
2012), via minor mergers between SGs and gas-rich dwarfs (Kaviraj
et al. 2009a; Kaviraj et al. 2011, see also Tal et al. 2012;
Newman et al. 2011).

While it is clear that the dominant stellar populations in SGs are
old, remarkably little is known about exactly when and how these
stars formed in the early Universe. The key questions are:
\emph{(1) At what redshifts and over what timescales did they
form? (2) What were the principal mechanisms that drove this star
formation?} While the former question has been explored using
spectro-photometric studies of local SGs, it is challenging to
construct a detailed star formation history for the old stars
using \emph{local} SGs, because the available spectro-photometric
indicators are insensitive to old stellar populations of different
ages. Optical colours, for example, evolve slowly after 4-5 Gyr
\citep[e.g.][]{Yi2003}. A further complication is that \emph{all}
indicators are affected by recent star formation, typically by an
uncertain amount in any given galaxy. While old stellar
populations do dominate SGs \emph{by mass}, measured
spectro-photometric quantities are \emph{luminosity-weighted}.
Since the luminosity-weighting of young stars is higher than their
old counterparts, even small amounts of recent star formation can
have a disproportionately large impact on the indicator in
question. Thus, while it is possible to put lower limits on the
age of the old stellar populations in today's SGs, the exact shape
of the star formation history of these old stars remains elusive.
It is worth noting that the \emph{average timescale} of star
formation in individual galaxies can be estimated with reasonable
precision using [$\alpha$/Fe] ratios\footnote{$\alpha$ elements
are primarily provided by prompt Type II supernovae, which explode
almost instantaneously ($\sim10^6$ yr) on cosmological timescales.
Fe, on the other hand, is provided by Type Ia supernovae (SN),
which emerge after typical time delays of less than a Gyr
\citep[e.g.][]{Thomas1999}. The [$\alpha$/Fe] ratio therefore
indicates the ratio of the overall timescale of star formation in
a galaxy to the onset timescale of the Type Ia SN.}. Indeed, the
super-solar [$\alpha$/Fe] ratios observed in most SGs suggest
star-formation timescales less than $\sim$ 1 Gyr
\citep{Trager2000a,Ferreras2000c,Thomas2005}.

\begin{figure}
\begin{center}
\includegraphics[width=3in]{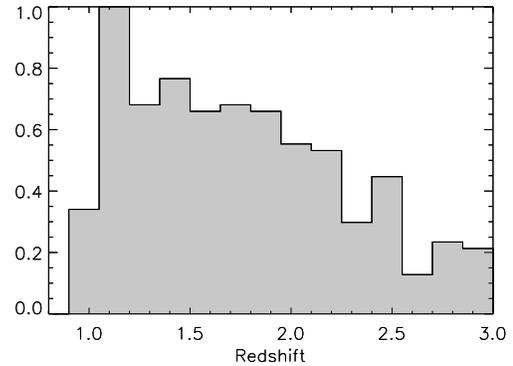}
\caption{Redshift distribution of the SGs in this study.}
\label{fig:rs_hist}
\end{center}
\end{figure}

\begin{figure}
\begin{center}
\includegraphics[width=3in]{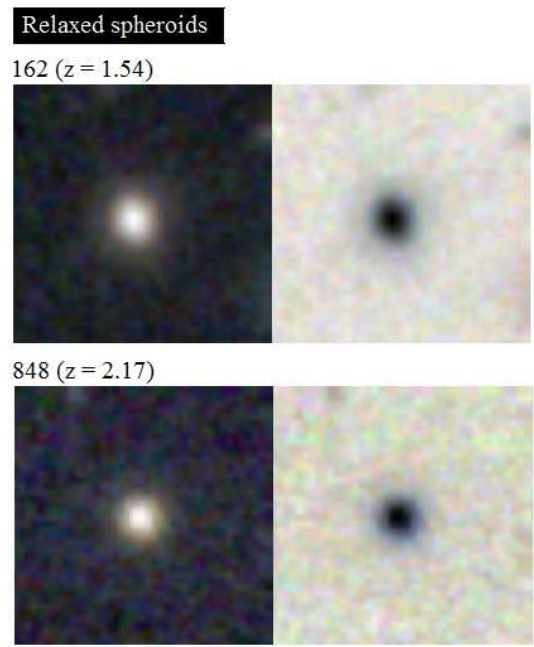}
\caption{Examples of relaxed SGs in our dataset. The redshift of
each galaxy is indicated above the image, next to its ID. We show
both the $YJH$ colour image and its negative.}
\label{fig:relaxed_SG_images}
\end{center}
\end{figure}

The latter question (2) has been the focus of some debate in the
literature. While the classical view has been that SGs are
products of `major' (roughly equal-mass) mergers between massive
spiral galaxies in the early Universe
\citep[e.g.][]{Toomre_mergers,White1978,Barnes1992a,Somerville1999,Cole2000,Hatton2003,Springel2005a},
recent hydrodynamical simulations
\citep[e.g.][]{Dekel2009a,Keres2009,Devriendt2010}, coupled with
the observed paucity of major mergers at high redshift
\citep[e.g.][]{Genzel2008,Law2009,Jogee2009,Kaviraj2011}, suggest
that the progenitors of SGs may not predominantly be major mergers
but rather clumpy disks, {\color{black}in which star formation is
fed by cold streams and minor mergers \citep[see
e.g.][]{Birnboim2003,Keres2005,Dekel2009b,Ceverino2010,Ceverino2012},
with the clumps eventually coalescing to form a spheroid.}

Given these significant open issues, a detailed \emph{empirical}
understanding of the formation of the dominant, old stars in
today's SGs is highly desirable, both as a route to understanding
the drivers of star formation in the early Universe, and as an
important test of our emerging theoretical models for the
high-redshift Universe. Given the limited utility of local SGs in
studying these issues, an ideal method for probing questions (1)
and (2) above is to study \emph{newborn} SGs {\color{black}around}
the epoch of peak cosmic star formation \citep[$1<z<3$,
see][]{Madau1998,Hopkins2004,Hopkins2006}. Quantifying the star
formation histories of these systems then allows us to directly
probe the {\color{black}formation} of the old stars in today's
SGs, free from the contamination of the galaxy spectrum by
intervening star formation episodes at late epochs. The underlying
assumption is that newborn SGs at high redshift are the ancestors
of their present-day counterparts, i.e. that SGs do not routinely
transform into disks over cosmic time. Both theoretical and
observational work on the evolving morphological mix of the
Universe suggests that this is indeed a reasonable assumption to
make \citep[e.g.][]{BO84,Baugh1996,Couch98,Smith2005}.

While the analysis of newborn SGs at $z>1$ is a compelling
project, {\color{black}past studies of statistically large samples
of SGs} \citep[e.g.][]{Ferreras2009} have typically fallen short
of this epoch, because deep, \emph{survey-scale} HST imaging in
the near-infrared (NIR), required for obtaining rest-frame optical
galaxy morphologies at $z>1$, has, until recently, been lacking.
However, following the GOODS-NICMOS survey \citep{Conselice2011},
a new generation of NIR surveys using the HST's WFC3 -- e.g. the
WFC3 Early-Release Science (ERS) programme \citep{Windhorst2011}
and the ongoing CANDELS campaign \citep{Koekemoer2011,Grogin2011}
-- are providing unprecedented large-scale access to rest-frame
optical galaxy data at $z>1$, making them ideal datasets for the
study of newborn SGs at high redshift. Here, we employ WFC3 ERS
data and study SGs around the epoch of peak cosmic star formation,
to explore the high-redshift origin of these galaxies and gain
{\color{black}insight} into when and how the old stellar
populations that dominate today's Universe formed.
{\color{black}We leverage the unprecedentedly deep NIR imaging to
morphologically identify SGs and explore the role of major mergers
in their evolution, and exploit the uniquely wide ERS wavelength
baseline (0.2 - 1.7 $\mu$m) to quantify galaxy star formation
histories via the available rest-frame UV-optical photometry.}

\begin{table*}
\begin{center}
\begin{minipage}{100mm}
\begin{tabular}{l|c|c|c|c}

    & \multicolumn{2}{|c|}{$1<z<1.5$} & \multicolumn{2}{|c|}{$1.5<z<3$}\\ \hline

    & f$_{\textnormal{SG}}$  & N$_{\textnormal{DSG}}$/N$_{\textnormal{SG}}$ &
      f$_{\textnormal{SG}}$  & N$_{\textnormal{DSG}}$/N$_{\textnormal{SG}}$\\ \hline

    Log M$_*>9.5$        & 0.35$^{\pm 0.02}$ & 0.42$^{\pm 0.04}$ & 0.32$^{\pm 0.02}$ & 0.35$^{\pm 0.04}$\\
    $9.5 < $Log M$_*<11$ & 0.36$^{\pm 0.03}$ & 0.43$^{\pm 0.05}$ & 0.36$^{\pm 0.02}$ & 0.36$^{\pm 0.04}$\\
    Log M$_*>11$         & 0.56$^{\pm 0.11}$ & 0.33$^{\pm 0.12}$ & 0.58$^{\pm 0.13}$ & 0.18$^{\pm 0.08}$

\end{tabular}
\end{minipage}
\caption{The fraction of galaxies that are SGs
(f$_{\textnormal{SG}}$) in various stellar mass (units of
M$_{\odot}$) and redshift ranges, and the fraction \emph{within}
the SGs that are disturbed
(N$_{\textnormal{DSG}}$/N$_{\textnormal{SG}}$). For the most
massive galaxies ($10^{11}$M$_{\odot}$-$10^{12}$M$_{\odot}$), for
example, the fraction of SGs is $\sim$60\%, similar to the value
found by Cameron et al. 2011, who also employed visual inspection
for identifying ETGs. The total number of galaxies in this study
is 818 (736 in the range $9.5 < $Log M$_*<11$ and 82 with Log
M$_*>11$).}
\end{center}
\end{table*}

\begin{figure}
\begin{center}
\includegraphics[width=3in]{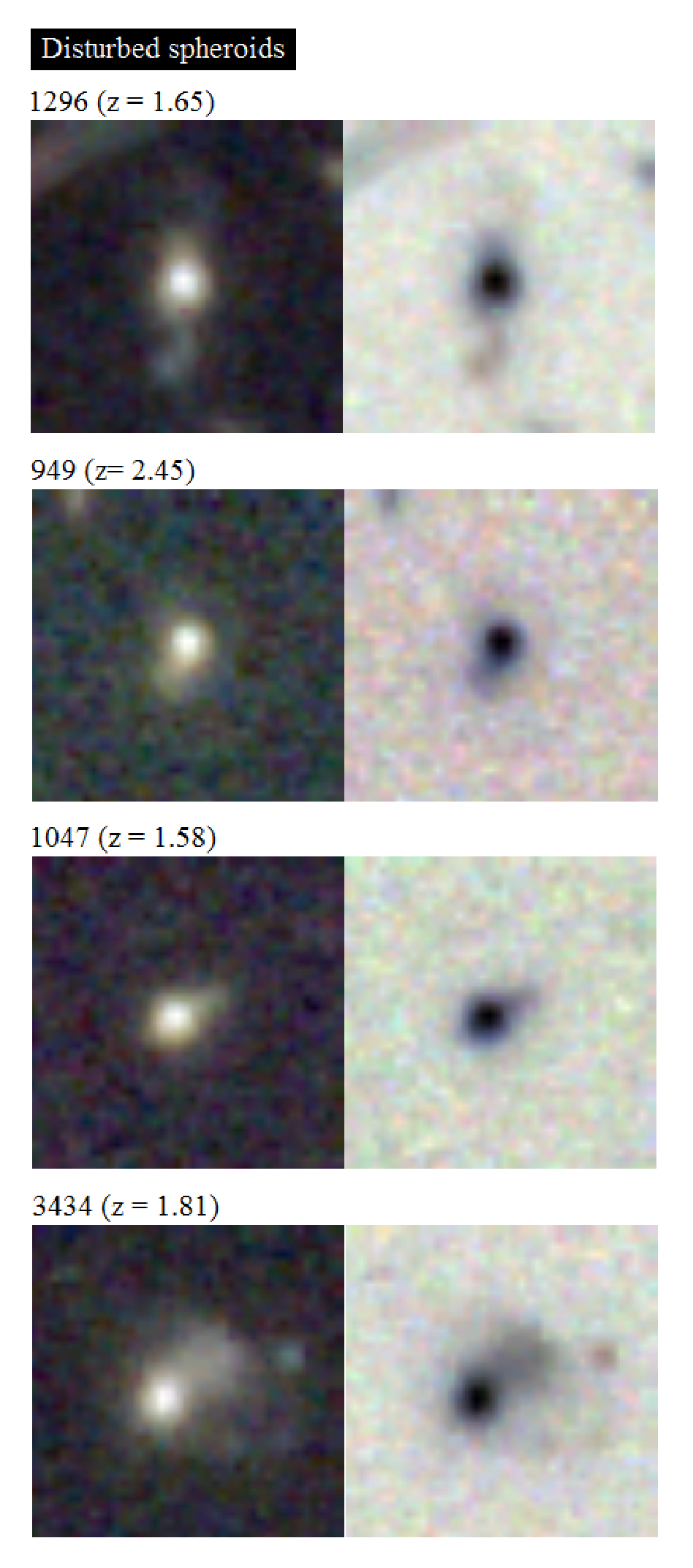}
\caption{Examples of disturbed SGs in our dataset. The redshift of
each galaxy is indicated above the image, next to its ID. We show
both the $YJH$ colour image and its negative.}
\label{fig:disturbed_SG_images}
\end{center}
\end{figure}

This paper is organised as follows. In Section 2, we briefly
describe the ERS dataset and the morphological selection of SGs
using visual inspection. In Section 3, we describe the calculation
of star formation histories and rest-frame photometry for
individual SGs in our sample. In Section 4, we study the
photometric properties of the newborn SGs, use their derived star
formation histories to {\color{black}constrain} the formation
epoch of their stellar mass and explore the mechanisms by which
this stellar mass is likely being formed. We summarise our
findings and outline avenues for future work in Section 5.
Throughout this paper we use the WMAP7 cosmological parameters
\citep{Komatsu2011}. {\color{black}All} photometry is presented in
the AB system \citep{Oke1983}.


\section{Galaxy sample and morphological classification}
{\color{black}The WFC3 ERS programme has imaged around one-third
of the GOODS-South field with both the UVIS and IR channels of the
HST/WFC3. The observations, data reduction, and instrument
performance are described in detail in Windhorst et al. (2011),
and highlighted here. The goal of this part of the ERS programme
was to demonstrate the science capabilities of the WFC3 for
studying intermediate- and high-redshift galaxies in both the UV
and NIR, by observing a portion of the well-studied GOODS-South
field \citep{Giavalisco2004}. The field was observed for 104
orbits, with 40 orbits in the UVIS channel and 60 orbits in the
NIR channel. The UVIS data covered $\sim$55 arcmin$^2$, in each of
the F225W, F275W and F336W filters, with relative exposure times
of 2:2:1. The IR data covered $\sim$45 arcmin$^2$ using the F098M
($Y_s$), F125W ($J$), and F160W ($H$) filters with equal exposure
times of 2 orbits per filter. The data were astrometrically
aligned with a version of the GOODS-S HST/ACS data
(v2.0\footnote{http://archive.stsci.edu/pub/hlsp/goods/v2/};
Giavalisco et al. 2004) that was rebinned to have a pixel scale of
$0\farcs09$ per pixel. Together, the data provide 10-band HST
panchromatic coverage over 0.2 - 1.7 $\mu$m, with a 5$\sigma$
point-source depth of $AB \sim 26.4$ mag in the UV and $AB
\sim27.5$ mag in the IR.

Our focus in this paper is the subset of 818 ERS galaxies that
have either spectroscopic or photometric redshifts in the range
$1<z<3$. Photometric redshifts are calculated using the EAZY code
\citep{Brammer2008} on the 10-band WFC3/ACS photometric catalogue
{\color{black}(Cohen et al. in prep)}. Spectroscopic redshifts are
drawn from the literature, from spectra taken using the Very Large
Telescope
\citep{LeFevre2004,Szokoly2004,Mignoli2005,Ravikumar2007,Vanzella2008,Popesso2009},
the Keck telescopes \citep{Strolger2004} and the HST ACS grism
\citep{Daddi2005,Pasquali2006,Ferreras2009}. For the analysis that
follows (e.g. in Section 3), spectroscopic redshifts are always
used where available}. {\color{black} Figure \ref{fig:rs_hist}
shows the redshift distribution of the SG sample in this study.}

Each galaxy in this sample is morphologically classified via
visual inspection of its WFC3/NIR images, which trace the
rest-frame optical wavelengths at $z>1$. While morphological
proxies, such as concentration, asymmetry, clumpiness, M$_{20}$
and the Gini coefficient have been widely used to classify
galaxies in large surveys \citep[see
e.g.][]{Abraham1996,Abraham2003,Conselice2003,Lotz2004}, the
performance of these methods is typically calibrated against
visual inspection \citep[e.g.][]{Abraham1996}, which offers better
precision and reliability in morphological classification
\citep[e.g.][]{Lisker2008,Robaina2009,Kartaltepe2010}.
{\color{black}Past} HST studies of the high-redshift Universe
(including recent efforts using the WFC3) have commonly exploited
visual inspection to classify galaxy morphologies
\citep[e.g.][]{Elmegreen2005,Ferreras2005,Bundy2005,Cassata2005,Jogee2009,Robaina2009,Kaviraj2011,Cameron2011,Kocevski2012},
using rest-frame optical imaging that has similar or fainter
surface-brightness limits compared to the images employed here.

\begin{figure}
\begin{center}
$\begin{array}{c}
\includegraphics[width=3.3in]{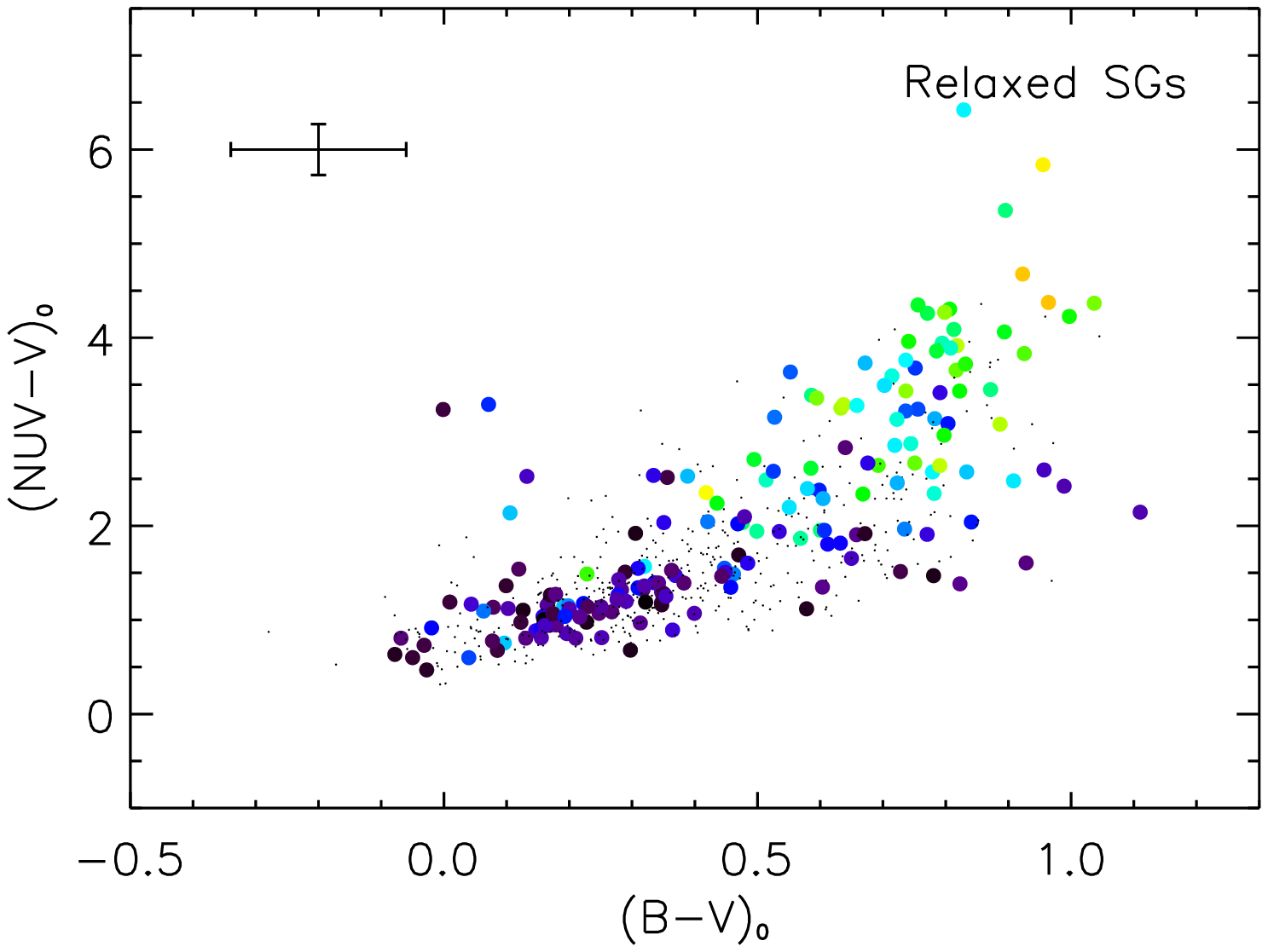}\\
\includegraphics[width=3.3in]{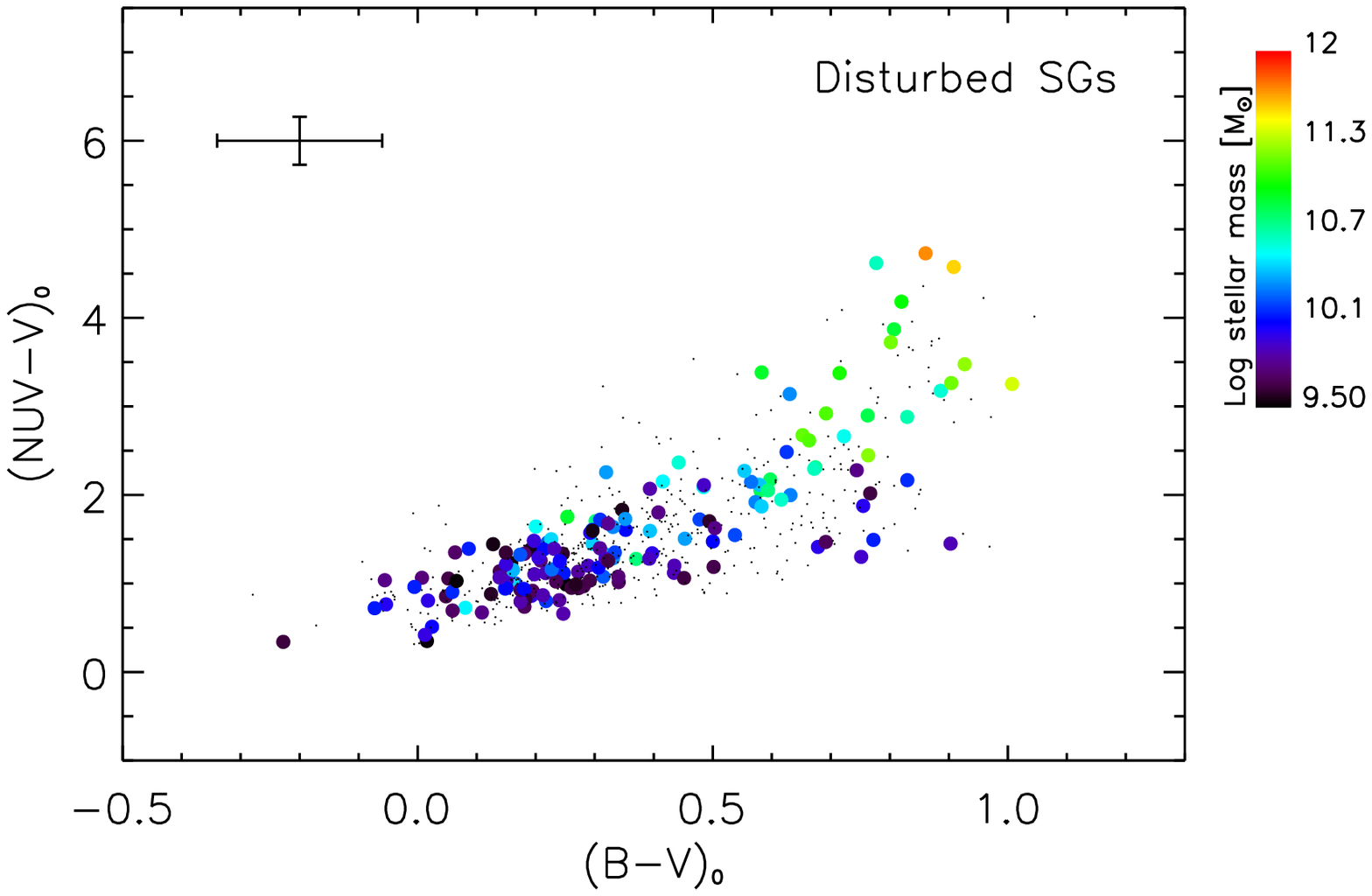}
\end{array}$
\caption{Rest-frame UV/optical colours of the ERS galaxies in the
redshift range $1<z<3$. The top panel shows SGs that are relaxed
i.e. do not show any morphological disturbances, while the bottom
panel shows SGs that are morphologically disturbed. The filled
coloured circles represent SGs, while the small black dots
represent late-type galaxies. The colour coding indicates the
stellar mass of the SG (see legend). The errors in the stellar
masses are typically better than 0.3 dex.}
\label{fig:rest_colours}
\end{center}
\end{figure}

\begin{figure}
\includegraphics[width=3.5in]{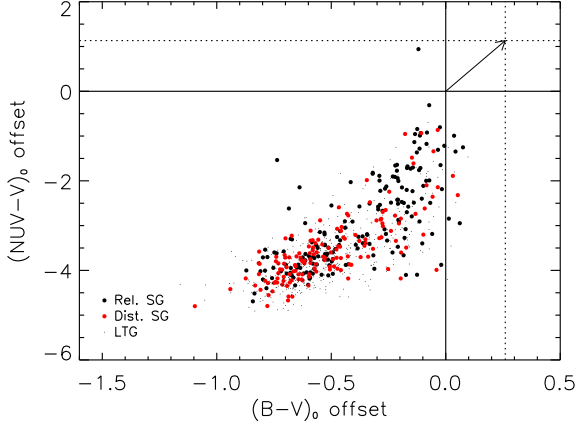}
\caption{The \emph{offset} between the rest-frame colours of
individual SGs and that of a dustless, solar-metallicity
instantaneous burst at $z=5$. The dotted line shows how the offset
will change if we consider a more realistic instantaneous burst
that is folded with the median value for the internal extinction
of the SGs ($E_{B-V} \sim 0.2$ mag). The arrow indicates the
reddening vector for this extinction value. {\color{black}Black
and red circles indicate relaxed and disturbed SGs respectively
and small black dots indicate late-type galaxies.} This
comparison, intended only as a guide, indicates that none of the
SGs in our sample are likely to have completed their stellar mass
assembly by $z\sim5$.} \label{fig:compare_to_z5}
\end{figure}

In this paper, we visually inspect $YJH$ composite images, scaled
using the asinh method of \citet{Lupton2004}, to determine (1) the
morphology of the galaxy and (2) whether it exhibits morphological
disturbances, i.e. tidal features indicative of a recent merger.
Using $YJH$ composites (instead of monochrome images in one
filter) maximises the rest-frame optical information in the image,
and facilitates the identification of tidal features. While
previous studies have successfully performed visual inspection of
similar datasets to $H\sim25$ mag \citep[e.g.][]{Cameron2011}, we
restrict our galaxy sample to a more conservative $H\sim24.5$ mag
(in this magnitude range the galaxy sample with M$_*>10^{9.5}$
M$_{\odot}$ is expected to be complete; Windhorst et al. 2011).
Galaxies are classified into two morphological classes: SGs and a
(broad) class of `late-types', which includes all objects that are
not spheroidal, e.g. disks, clumpy galaxies, irregulars, etc.
Here, we focus solely on the SGs, splitting them into `relaxed'
systems, which do not exhibit morphological disturbances
indicative of recent interactions, and `disturbed' systems, which
do (see Figures \ref{fig:relaxed_SG_images} and
\ref{fig:disturbed_SG_images} for examples of relaxed and
disturbed SGs respectively). {\color{black} The visual
classification process identifies 328 SGs in our sample.} As we
describe in Section 4 below, the disturbed SGs are likely to have
experienced recent \emph{major} mergers.

For the analysis that follows, we study ERS galaxies that have
redshifts in the range $1<z<3$, $H$-band magnitudes brighter than
24.5 (where morphologies are reliable) and stellar masses greater
than $10^{9.5}$ M$_{\odot}$ {\color{black}(the derivation of
stellar masses is described in Section 3 below)}. Table 1
summarises the fraction of galaxies that are SGs
(f$_{\textnormal{SG}}$) in various stellar mass and redshift
ranges and the fraction \emph{within} the SGs that are disturbed
(N$_{\textnormal{DSG}}$/N$_{\textnormal{SG}}$). For example, for
the most massive galaxies (M$_*>10^{11}$ M$_{\odot}$), the
fraction of SGs is $\sim$60\%, similar to the value found by
Cameron et al. (2011), who also used visual inspection of WFC3
images for morphological classification. In the mass range
considered here (M$_*>10^{9.5}$ M$_{\odot}$), around a third of
SGs exhibit morphological disturbances. The disturbed fraction
decreases with increasing redshift, plausibly due to greater
cosmological dimming of the tidal features at larger distances.


\section{Star formation histories and rest-frame photometry}
In this section, we describe the estimation of star formation
histories (SFHs) and rest-frame photometry that are used in
Section 4. We compare the observed photometry of each galaxy to a
large library of synthetic photometry. The synthetic library is
constructed using exponentially-decaying model SFHs, described by
five free parameters: the age ($T$), {\color{black}e-folding}
timescale ($\tau$), metallicity ($Z$) and internal extinction
($E_{B-V}$) of the starburst and the total stellar mass formed
({\color{black}$M_*$}). We vary {\color{black}$T$} between 0.05
Gyrs and the look-back time to $z=20$ in the rest-frame of the
galaxy, $\tau$ between 0.01 Gyrs (approximately an instantaneous
burst) and 9 Gyrs (approximately constant star formation), $Z$
between 0.05 Z$_{\odot}$ and 2.5 Z$_{\odot}$ and $E_{B-V}$ between
0 and 2 mag. Synthetic photometry is generated for each model SFH
by {\color{black}folding} it with the stellar models of Charlot \&
Bruzual (2007) (an updated version of the \citet{BC2003} models),
through the correct WFC3 and ACS filter throughputs. The empirical
law of \citet{Calzetti2000} is used to compute the dust-extincted
spectral energy distributions (SEDs). The synthetic library has
$\sim2.5$ million individual models. Since our galaxies span a
large range in redshift ($1<z<3$), equivalent libraries are
constructed at redshift intervals of $\delta z=0.01$.

\begin{figure*}
\begin{minipage}{172mm}
\begin{center}
\includegraphics[width=\textwidth]{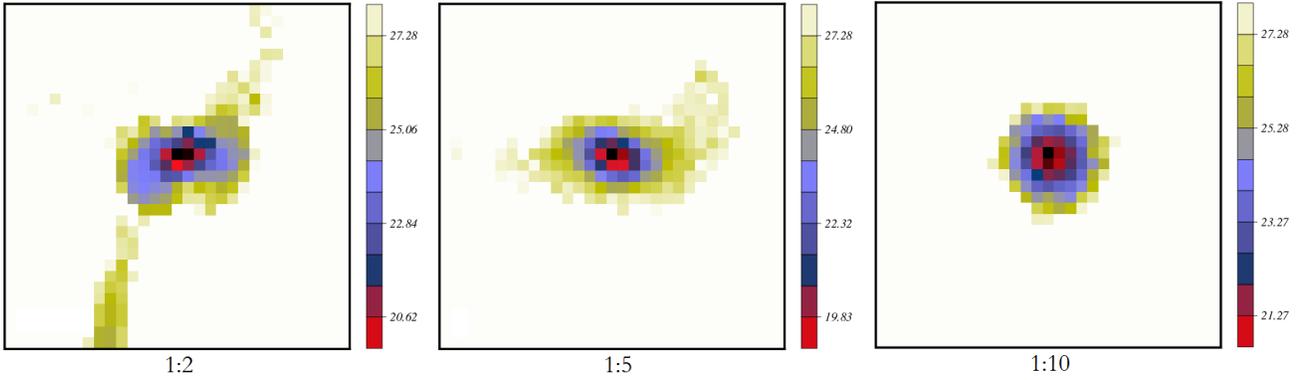}
\caption{Typical remnants of mergers, drawn from a hydrodynamical
cosmological simulation, with mass ratios of 1:2 (left), 1:5
(middle) and 1:10 (right). The colour-coding indicates the
surface-brightness (mag arcsec$^{-2}$) in the $H$-band (the
surface brightness limit of our WFC3/IR images is $\sim$ 26 mag
arcsec$^{-2}$). To allow comparison with the blue SGs in our
sample, the remnants are `observed' while they are in the
UV-optical blue cloud, typically $\sim$0.3-0.4 Gyr after the
merger is complete. The simulations suggest that tidal debris
around remnants of mergers that have mass ratios less than
$\sim$1:5 are very unlikely to be observable at the depth of the
ERS images across the redshift range considered in this study. In
other words, blue SGs that do exhibit such tidal features are
likely to be \emph{major-merger remnants} (see text for more
details).} \label{fig:simulated_merger_examples}
\end{center}
\end{minipage}
\end{figure*}

For each galaxy, the values of the free parameters ($T$, $\tau$,
{\color{black}$M_*$}, $Z$, $E_{B-V}$) are estimated by comparing
its observed photometry to every model in the synthetic library
that is closest to it in redshift. The likelihood of each model,
$\exp (-\chi^2/2)$, is calculated using the value of $\chi^2$,
computed in the standard way. From the joint probability
distribution (which is a function of all five free parameters),
each individual parameter is marginalised\footnote{The
one-dimensional marginalised probability distribution (PD) for age
($T$), for instance, is obtained from the joint PD by integrating
out the other parameters. If $\textnormal{P} (\textnormal{T},
\tau, \textnormal{M}, \textnormal{Z},
\textnormal{E}_{\textnormal{B-V}}| \textnormal{D})$ is the joint
PD (given the data $\textnormal{D}$), then the marginalised PD for
$\textnormal{T}$,
$\textnormal{P}(\textnormal{T}|\textnormal{D})=\int \int \int
\int^{\infty}_0 \textnormal{P}(\textnormal{T}, \tau,
\textnormal{{\color{black}M$_*$}}, \textnormal{Z},
\textnormal{E}_{\textnormal{B-V}}|\textnormal{D}) \hspace{0.05in}
\textnormal{d}\tau \hspace{0.05in}
\textnormal{{\color{black}dM$_*$}} \hspace{0.05in} \textnormal{dZ}
\hspace{0.05in}
\textnormal{dE}_{\textnormal{B-V}}$.\vspace{0.05in}} to extract
its one-dimensional probability density function (PDF). The median
of this PDF is taken as the best estimate of the parameter in
question and the 25 and 75 percentile values {\color{black}(which
enclose 50\% of the probability)} are used to calculate an
uncertainty on this estimate. The K-corrections required to
construct rest-frame photometry for each galaxy are calculated
using the best-fit model SED (i.e. where the value of $\chi^2$ is
a minimum).

It is worth noting that the accuracy of the derived star formation
histories is aided by the availability of photometry covering both
the rest-frame UV and optical wavelengths. The rest-frame UV, in
particular, offers almost an order of magnitude greater
sensitivity to recent star formation {\color{black}than} the
optical wavelengths (Kaviraj et al. 2007a), and is largely free
from the effects of the age-metallicity degeneracy (Kaviraj et al.
2007b). The rest-frame UV colours are therefore a strong indicator
of how {\color{black}truly} quiescent a galaxy is. While the
rest-frame UV photometry constrains the gradual quenching/decline
of star formation in our SGs, the rest-frame optical constrains
the epoch of stellar mass assembly.

\section{The mass assembly of newborn spheroids}
We begin by exploring the photometric properties of our ERS
dataset. {\color{black}Figure \ref{fig:rest_colours} presents the
\emph{rest-frame} UV-optical colours of the ERS galaxies. The
rest-frame $NUV$ wavelengths, based on the GALEX
\citep{Martin2005} $NUV$ filter, are centred at $\sim$2300 \AA}.
Relaxed and disturbed SGs span the entire colour space occupied by
the galaxy population, with $\sim$40\% and $\sim$60\% of relaxed
and disturbed SGs respectively populating the `blue cloud'
($B-V<0.5$ and $NUV-V<2.5$). This is consistent with the large
range in rest-frame ($UVJ$) optical colours observed in the
massive galaxy population at these redshifts
\citep[e.g.][]{Whitaker2010} and the gradual, rather than abrupt,
decline of star formation implied in most massive objects in this
redshift range by their spectral features
\citep[e.g.][]{Cimatti2008,VD2010,Kriek2011,VD2011}. In Figure
\ref{fig:compare_to_z5} we show the \emph{offset} between the
rest-frame colour of individual SGs and that of a dustless,
solar-metallicity instantaneous burst at $z=5$. This comparison,
intended only as a guide, indicates that very few of the SGs are
consistent with having completed their star formation by $z\sim5$.
Note that, adding the median internal extinction in SGs derived in
Section 3 ($E_{B-V} \sim 0.2$ {\color{black}mag}, see reddening
vector in Figure \ref{fig:compare_to_z5}) to this instantaneous
burst, only reinforces this conclusion. \emph{The star formation
in SGs at 1 $<$ z $<$ 3 is, therefore, either ongoing or recently
completed.}

It is worth noting that around half of the SGs that inhabit the
blue cloud appear relaxed. {\color{black}Given that tidal features
\emph{are} readily visible in a significant fraction of objects,
we explore whether the lack of such features in the \emph{blue
relaxed} SGs implies that the star formation in these systems is
not being driven by a recent major merger}. Indeed, tidal debris
from a recent merger is most readily visible in the early stages
of relaxation, when star formation remains strong and the galaxy
is in the blue cloud \citep[e.g.][]{Carpineti2012}.

\begin{figure}
\includegraphics[width=3.5in]{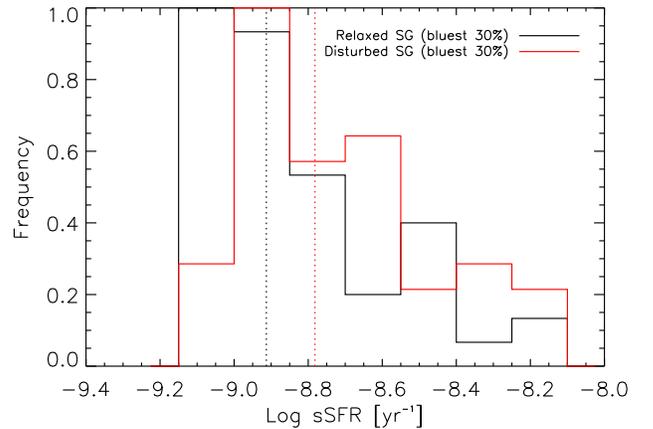}
\caption{Specific star formation rates (sSFRs) in the bluest 30\%
of relaxed (black) and disturbed (red) SGs {\color{black}(in the
$NUV-r$ colour)}. Median values are shown using the dotted lines.
The general enhancement in sSFR due to a merger appears to be
modest, with the median sSFR of the bluest 30\% of disturbed SGs
being around 0.15 dex ($\sim$40\%) higher than the corresponding
median in their relaxed counterparts. Median values are shown
{\color{black}by} the dotted vertical lines. {\color{black}Note
that the SFR is an average value over the last $10^7$ yr.}}
\label{fig:ssfr}
\end{figure}

\begin{figure*}
\begin{minipage}{172mm}
\begin{center}
$\begin{array}{cc}
\includegraphics[width=3.5in]{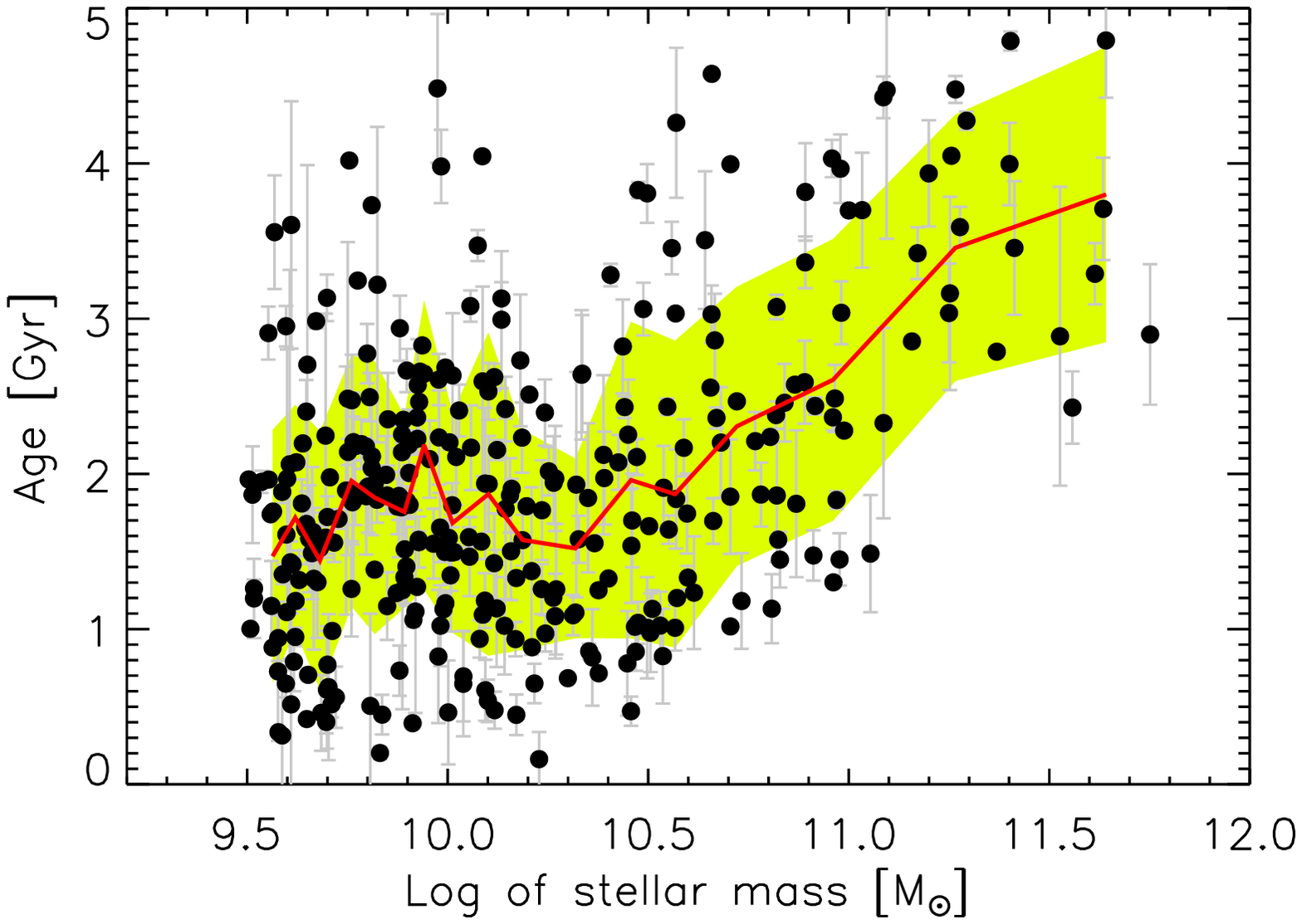} & \includegraphics[width=3.5in]{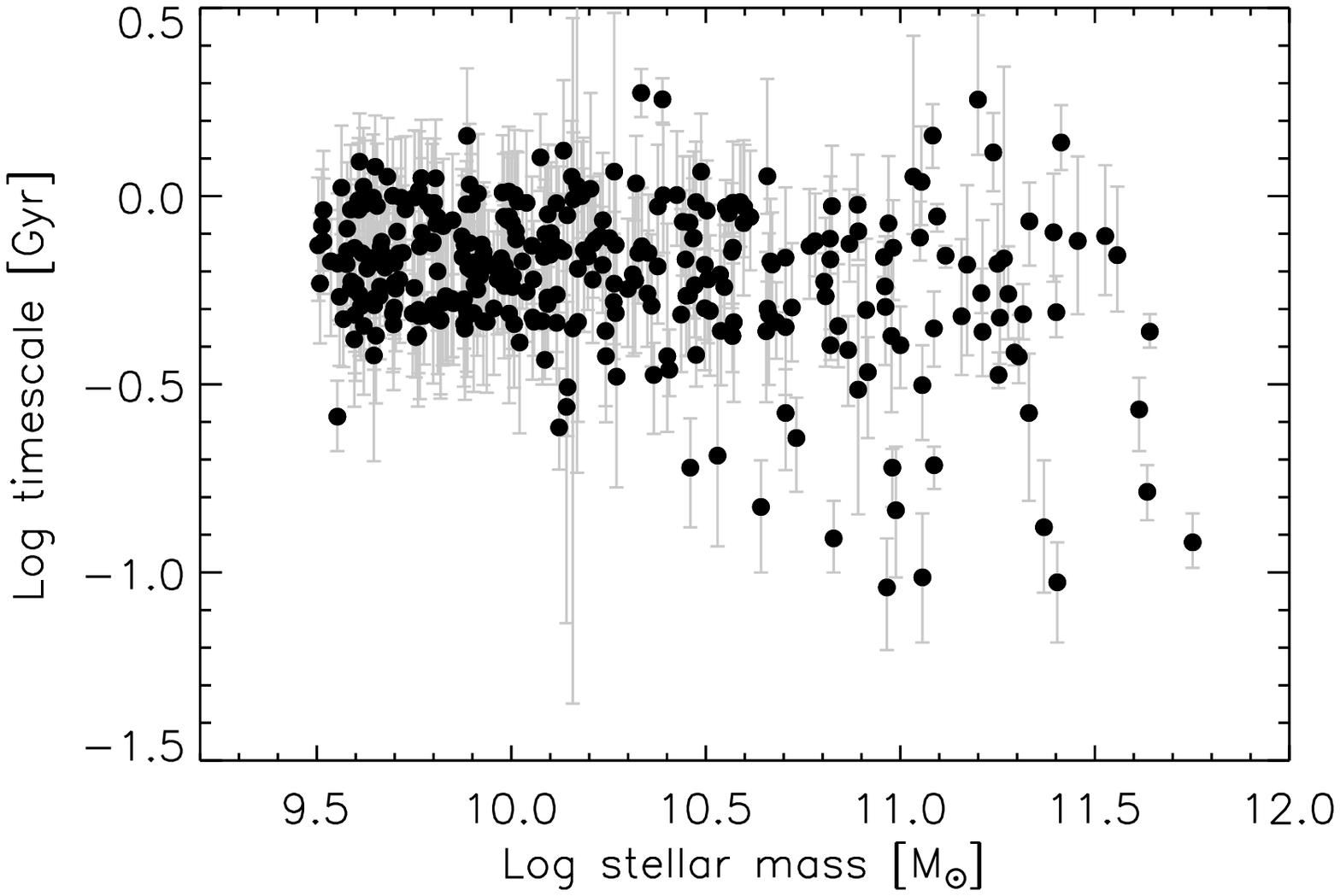}\\
\includegraphics[width=3.5in]{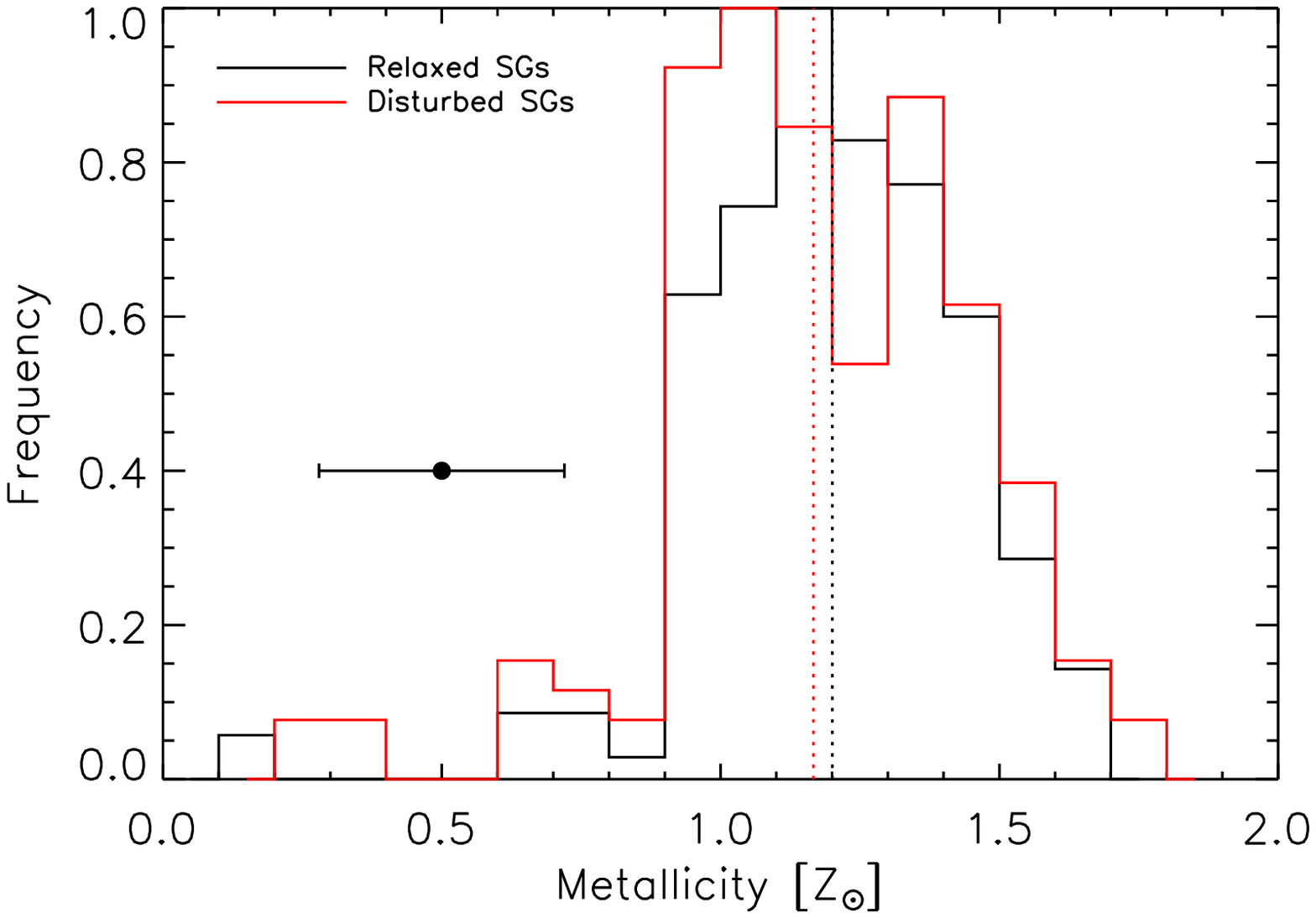} & \includegraphics[width=3.5in]{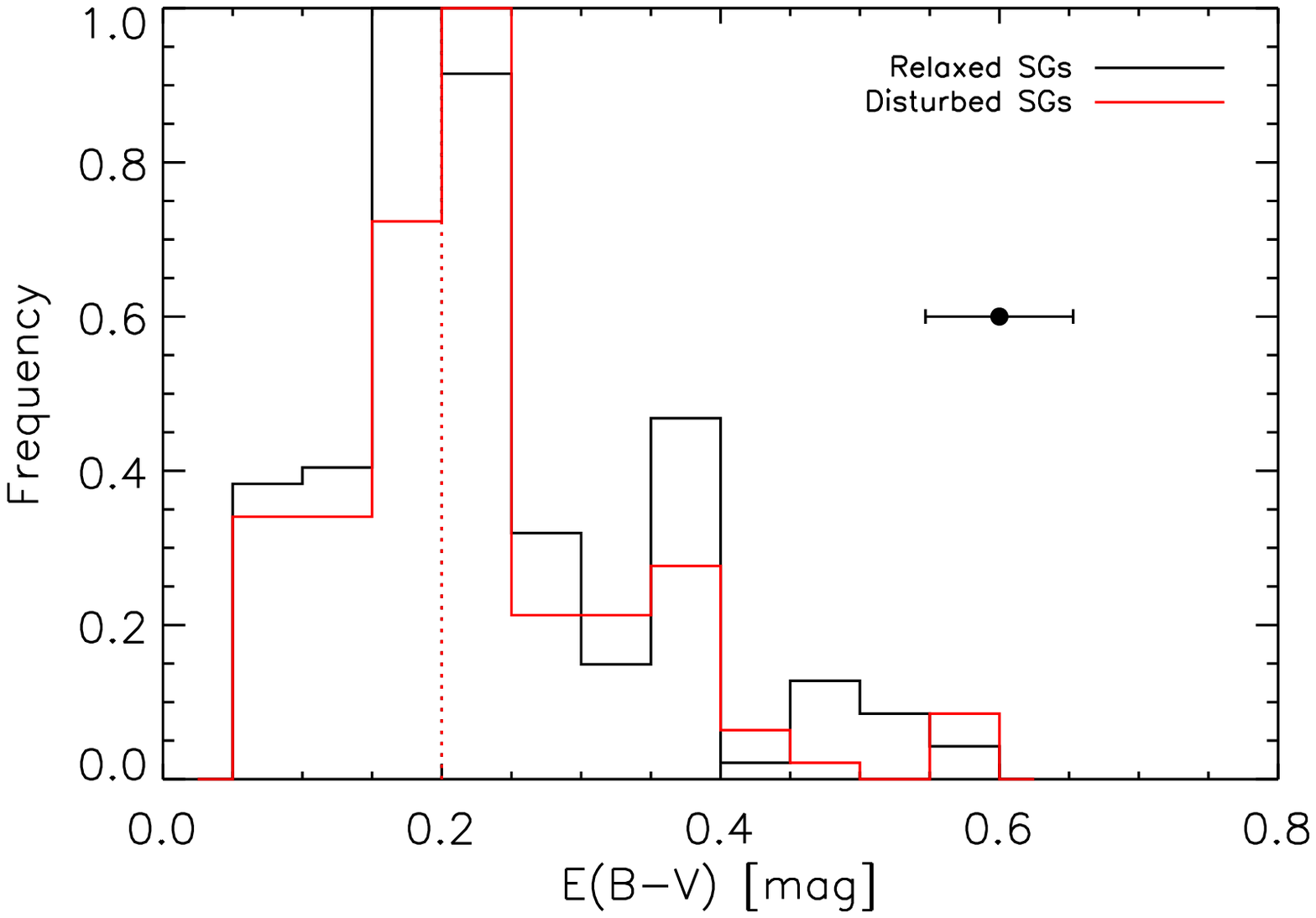}
\end{array}$
\caption{TOP LEFT: Derived starburst age {\color{black}($T$)} vs.
stellar mass in individual SGs. The yellow shaded region in the
top-left panel shows a progressive `one-sigma' fit to the data.
The red line indicates the mean value, while the yellow shaded
area indicates the region that encloses 68\% of the galaxies. The
typical errors in the stellar masses are better than 0.3 dex. TOP
RIGHT: Derived starburst timescale {\color{black}($\tau$)} vs
stellar mass in individual SGs. BOTTOM LEFT: Distribution of
stellar metallicities in individual SGs. BOTTOM RIGHT:
Distribution of internal extinction in individual SGs. Median
values are shown using the dotted vertical lines.
{\color{black}The error bars in the panels for metallicity and
internal extinction indicate average uncertainties in the derived
parameters across our SGs.}} \label{fig:sfh_parameters}
\end{center}
\end{minipage}
\end{figure*}

{\color{black}To explore the detectability of tidal features in
our high-redshift SGs, we appeal to merger remnants of various
mass ratios drawn from a hydrodynamical cosmological simulation.
The simulation, which is described in detail in
\citet{Peirani2010a}, was performed using the GADGET2 code
\citep{Springel2005b} with added prescriptions for star formation,
feedback from Type Ia and II supernovae, a UV background and metal
enrichment. The dark matter and baryonic particle resolutions are
$m_{DM} = 7.4 \times 10^6 h^{-1}$ M$_{\odot}$ and $m_{gas} =
m_{star} = 1.5 \times 10^6 h^{-1}$ M$_{\odot}$ respectively. We
refer readers to Section 2.1 of \citet{Peirani2010a} for further
details of the simulation.

To compare with the blue SGs in our sample, we explore the surface
brightness of tidal features in merger remnants at $z\sim1.25$,
the midpoint of the lower redshift bin in Table 1. We focus on the
lower redshift bin in order to minimise the impact of cosmological
dimming - recall that the disturbed SG fraction decreases with
redshift in Table 1, plausibly due to cosmological dimming of the
tidal features with increasing redshift. In Figure
\ref{fig:simulated_merger_examples} we present typical examples of
remnants of mergers that have ratios of 1:2, 1:5 and 1:10. The
merger remnants are `observed' while they are in the UV-optical
blue cloud (to ensure consistency with our blue observed SGs) and
the synthetic images are constructed by combining the simulation
outputs with the correct WFC3 filter throughputs, following
\citet{Peirani2010b}. Note that our aim here is to study the
surface brightness of \emph{tidal features} around merger remnants
and not a detailed exploration of the morphology of the remnants
themselves, which would benefit from a simulation with higher
resolution.

We find that, while major mergers produce strong features which
will be readily visible, mergers with mass ratios of 1:5 are only
marginally detectable at $z \sim 1.25$, given the
surface-brightness limit of extended objects in our WFC3/IR images
($\sim$ 26 mag arcsec$^{-2}$). Mergers with mass ratios lower than
this value (see e.g 1:10 example in Figure
\ref{fig:simulated_merger_examples}) will produce tidal features
that will not be visible in our images. At the upper limit of our
redshift range (z$\sim$3), only mergers with mass ratio $\sim$1:2
will be visible.  \emph{In summary, it is likely that only tidal
features produced by major mergers are likely to be visible across
the redshift range considered in this study.} In other words, if a
\emph{blue} SG in our high-redshift sample has experienced a
recent \emph{major} merger, then the tidal debris should be
visible {\color{black}in the ERS images}. The dominant star
formation mechanism in the blue relaxed SGs therefore appears
unlikely to be a major merger, implying that these systems are
driving their star formation via other mechanisms, such as minor
mergers, or are the remnants of recently collapsed
(cold-stream-fed) clumpy disks, as envisaged in current
theoretical work \citep[e.g.][]{Dekel2009a,Elmegreen2008}.}
{\color{black}Note that minor-merger-driven star formation could,
in principle, be distinguished from disk collapse by the presence
of tidal features. However, as Figure
\ref{fig:simulated_merger_examples} indicates, our WFC3 images are
too shallow to detect tidal debris from minor mergers. While it is
challenging to disentangle these two processes using current data,
it might be possible to perform this exercise using deeper data,
either using the WFC3 itself or from future instruments such as
the JWST or the extremely large telescopes.}

It is worth noting further that the relaxed SGs in the blue cloud
have a similar UV-optical colour distribution in Figure
\ref{fig:rest_colours} {\color{black}as} their disturbed
counterparts. Since colours reflect the (average) specific star
formation rate (sSFR), this suggests that the sSFR enhancement due
to a recent major merger is modest. Figure \ref{fig:ssfr} shows
that the median sSFR of the bluest 30\% {\color{black}(in the
$NUV-r$ colour)} of disturbed SGs is around 0.15 dex ($\sim$40\%)
higher than the corresponding median value in their relaxed
counterparts. A similarly modest enhancement in sSFR ($\leq$60\%)
due to major mergers has been reported in recent cosmological
simulations \citep{Cen2011} within similar stellar mass and
redshift ranges as those studied in this paper. {\color{black}Note
that the SFR is an average value over the last $10^7$ yr.}
\emph{Major mergers thus appear relatively insignificant, both in
terms of driving the buildup of the SG stellar mass, and enhancing
the star formation that is already being driven by other processes
e.g. cold-mode accretion and minor mergers.} {\color{black}While
our approach is empirical in nature, our conclusions regarding the
overall insignificance of major mergers appears consistent with
both recent theoretical work \citep[e.g.][]{Cen2011,Dekel2009a}}
and an emerging observational literature that indicates a high
fraction of systems with disk-like properties
\citep[e.g.][]{Shapiro2008,Stockton2008,ForsterSchreiber2006} and
a remarkably modest incidence of major mergers \citep[see
e.g.][and references therein]{Genzel2008,Tacconi2010} amongst
star-forming galaxies at $z \sim2$.

We conclude by summarising the derived star-formation histories of
individual SGs. In the top row of Figure \ref{fig:sfh_parameters},
we plot the starburst ages {\color{black}($T$)} and timescales
{\color{black}($\tau$)} of individual SGs against their stellar
masses. In the bottom row in this figure we show the distributions
of metallicities and internal dust extinctions for our SGs. The
yellow shaded region in the top-left panel shows a progressive
`one-sigma' fit to the data. The red line indicates the mean
value, while the yellow shaded area indicates the region that
encloses 68\% of the galaxies. Individual error bars for each SG
are shown and the errors in the galaxy stellar masses are
typically better than 0.3 dex.

\begin{figure}
\begin{center}
$\begin{array}{c}
\includegraphics[width=3.5in]{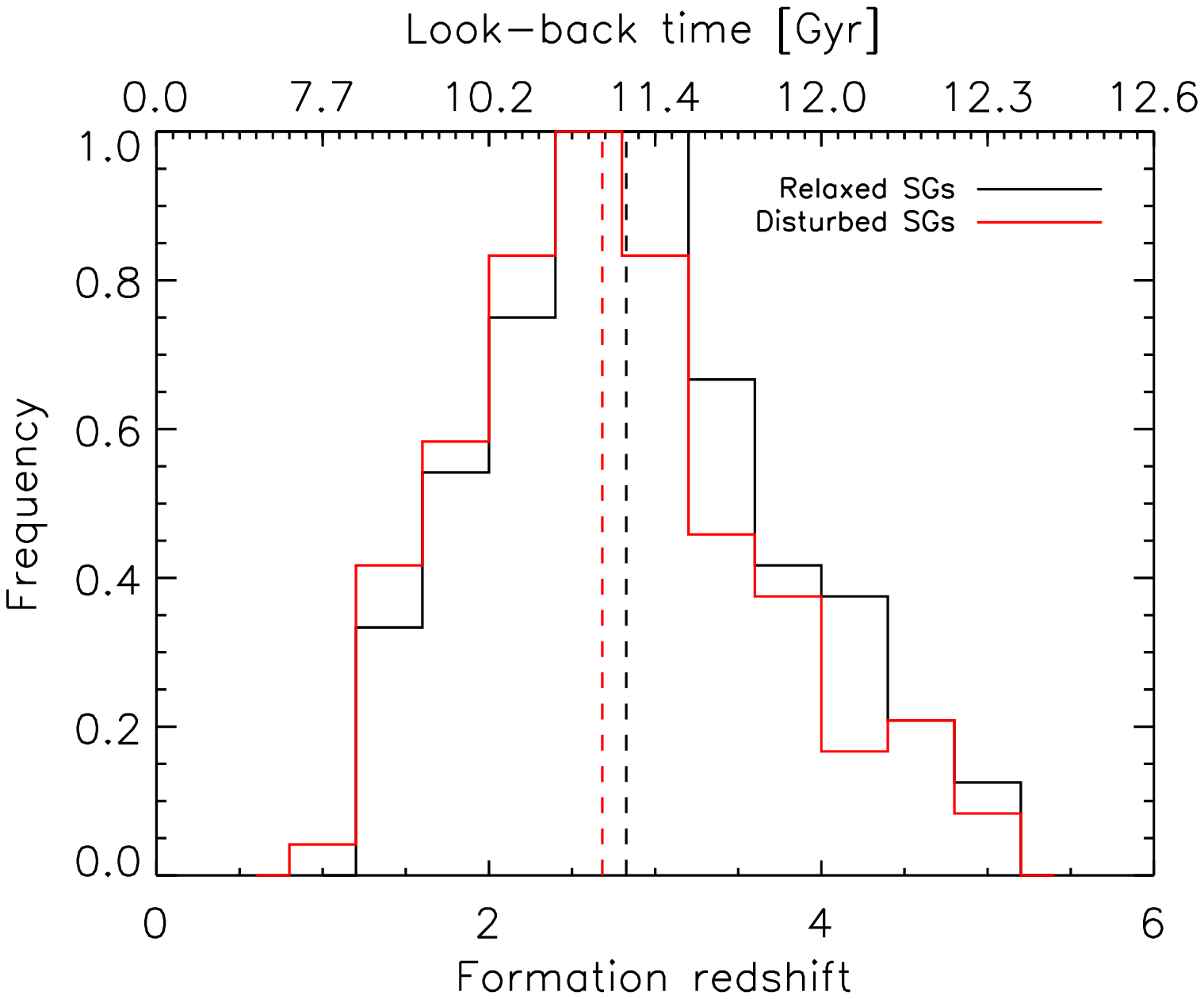}\\
\includegraphics[width=3.5in]{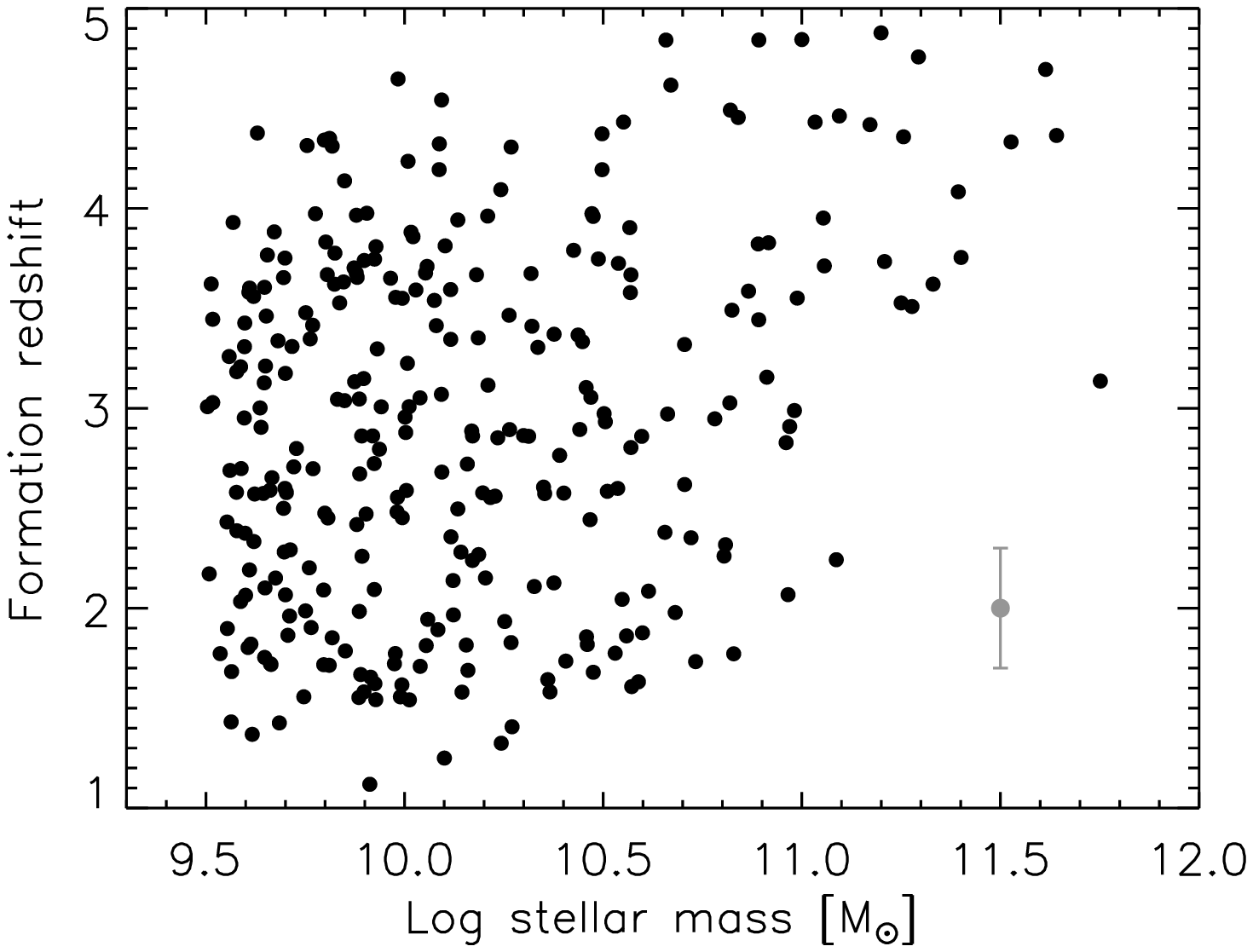}
\end{array}$
\caption{TOP: Histogram of formation redshifts, calculated from
the starburst ages {\color{black}($T$)} and the observed redshifts
of the SG population (see text in Section 4 for details). Median
values are shown using the dotted vertical lines. {\color{black}
Look-back times corresponding to the formation redshifts are
indicated on the top x-axis. BOTTOM: Formation redshift vs stellar
mass in individual SGs. The error bar indicates the average
uncertainty in the formation redshifts across the SG population.
The typical errors in the stellar masses are better than 0.3
dex.}} \label{fig:zform}
\end{center}
\end{figure}

We find that the star-formation timescales {\color{black}($\tau$)}
in SGs are relatively short, with a median of $\sim$0.6 Gyr, with
uncertainties on individual timescales of $\sim$0.3-0.4 dex.
{\color{black}Given the large uncertainties, timescales as large
as $\sim$1.5 Gyr cannot be ruled out for most of the individual
SGs in our sample. However, the median value for the sample as a
whole is smaller than 1 Gyr, consistent with previous measurements
of star formation timescales in intermediate and high-redshift SGs
\citep[e.g.][]{Ferreras2004} and the high values of alpha
enhancement found in \emph{local} SGs \citep[e.g][]{Thomas2005}}.

In the mass range $10^{9.5}$M$_{\odot}<$ M$_*<10^{10.5}$
M$_{\odot}$, the starburst timescales {\color{black}($\tau$)} and
ages {\color{black}($T$)} do not correlate strongly with galaxy
stellar mass. However, the timescales scatter towards lower values
($<0.3$ Gyr) for more massive galaxies
(M$_*>10^{10.5}$M$_{\odot}$) and {\color{black}a trend towards
increasing age} is apparent as we move to systems with higher
stellar masses. For example, galaxies at the upper end of our mass
range (M$_*>10^{11.5}$M$_{\odot}$) are typically $\sim$2 Gyrs
older than their counterparts with masses
$\sim10^{10.5}$M$_{\odot}$. {\color{black}These trends are
qualitatively consistent with both observational
\citep[e.g.][]{Thomas2005,Juneau2005} and theoretical evidence
\citep[e.g.][]{Neistein2006,Cattaneo2008} for `downsizing' in
massive galaxies. It is worth noting, however, that a
\emph{smooth} downsizing trend with galaxy mass is not observed,
and the large scatter in the starburst ages {\color{black}($T$)}
indicate that SGs are not a particularly coeval population. While
such coevality has been suggested in the past by the optical
colours of local SGs, this is likely due to the fact that 8-10 Gyr
of evolution washes out the \emph{details} of the stellar mass
assembly. One also has to entertain the possibility that the
trends observed in local SGs, between stellar mass and quantities
such as luminosity-weighted age, are influenced by late star
formation. For example, an identical gas-rich satellite will
likely create a larger mass fraction of young stars, and therefore
a lower luminosity-weighted age and a larger dilution of
[$\alpha$/Fe], in a smaller SG.}

The median value for the internal dust extinction of our SG sample
is $E_{B-V} \sim 0.2$ mag, around a factor of 2 higher than the
typical values found in local SGs (Kaviraj et al. 2007a).
{\color{black}Super-solar metallicities ($Z>Z_{\odot}$) are
favoured}, but with significant uncertainties that are typical of
this type of photometric parameter estimation. We conclude this
section by presenting the `formation redshifts' of the SG
population in Figure \ref{fig:zform}. The formation redshift is
calculated using the derived starburst age and observed redshift
of the SG in question , i.e. it is the redshift around which the
{\color{black}star formation} in the galaxy is likely to peak in a
WMAP-7 cosmology. \emph{The top panel in Figure 8 indicates that
the peak of the star formation that builds the SGs
{\color{black}in our sample} is likely in the redshift range
$2<z<5$, with a median value of $z\sim3$.} {\color{black}The
bottom panel in this figure shows the formation redshifts plotted
against the stellar masses for our SGs. The general trends are
similar to the plot of age ($T$) vs stellar mass (top-left panel
of Figure \ref{fig:sfh_parameters}), with the larger scatter
mainly due to the non-linearity between look-back time and
redshift.}


\section{Summary}
We have studied $\sim$330 newborn spheroidal galaxies (SGs) in the
redshift range $1<z<3$, to study their high-redshift origin and
gain insight into when and how the old stellar populations that
dominate today's Universe formed. SGs have been identified by
visual inspection of near-infrared images from the WFC3
Early-Release Science programme, which trace the rest-frame
optical wavelengths at these redshifts. Ten-filter HST photometry,
covering the rest-frame UV-optical wavelengths, has been used to
study the photometric properties of the newborn SG population and
empirically estimate their star formation histories.

The rest-frame UV-optical colours of the SG population indicate
that virtually none of these galaxies has completed their stellar
assembly by $z\sim5$. {\color{black}The derived star formation
histories indicate that the stellar assembly of our SGs likely
peaked in the redshift range $2<z<5$, with a median value of
$z\sim3$. Given that around half the present-day SG population was
in place by $z\sim1$ \citep[e.g.][]{Bell2005,Faber2007}, this
implies that a significant fraction of the old stars that dominate
the local Universe are likely to have formed at these epochs.}

Our results show that the star formation episodes that built the
massive SGs are relatively short and have decay timescales less
than 1.5 Gyr (with a median of $\sim$0.6 Gyr). Starburst ages and
timescales show no correlation with galaxy stellar mass in the
mass range $10^{9.5}$M$_{\odot}<$ M$_*<10^{10.5}$ M$_{\odot}$.
However, the timescales scatter towards lower values ($<0.3$ Gyr)
for more massive galaxies (M$_*>10^{10.5}$M$_{\odot}$) and
{\color{black}a trend towards increasing age} becomes apparent as
we move to higher stellar masses, with galaxies that have
M$_*>10^{11.5}$M$_{\odot}$ being $\sim$2 Gyrs older than those
with M$_*<10^{10.5}$M$_{\odot}$. However, a \emph{smooth}
downsizing trend with galaxy mass is not observed, and the large
scatter in the starburst ages indicate that SGs are not a
particularly coeval population.

Around half of the SGs in the blue cloud are \emph{relaxed}, i.e.
show no morphological disturbances of a recent merger. At the
depth of the WFC3 images employed in this study, tidal debris from
\emph{major} mergers is likely to be visible at the epochs probed
here. Thus, relaxed SGs in the blue cloud are unlikely to be
driving their star-formation episodes via major mergers,
suggesting (indirectly) that they may be experiencing minor
mergers, or are the remnants of the recent collapse of clumpy
disks, in which the star formation has been fed by cold streams
(as envisaged in recent theoretical work). Furthermore, those SGs
that do show tidal features, and are therefore likely to be recent
major-merger remnants, exhibit only modest enhancements in their
specific star formation rates of $\sim$40\%. {\color{black}Thus,
major mergers appear relatively insignificant, both in terms of
driving the buildup of the SG stellar mass and enhancing the star
formation that is already being driven by other processes (e.g.
cold-mode accretion, minor mergers.)}

{\color{black}This study offers \emph{empirical} insights} into
the {\color{black}formation} of newborn SGs in the early Universe,
using the rest-frame optical data that is rapidly becoming
available from new space and ground-based surveys. We conclude
this paper by outlining several outstanding issues that demand
further study. The analysis of very massive galaxies
(M$_*>10^{11}$M$_{\odot}$) requires a larger,
statistically-significant sample of objects, such as the complete
CANDELS survey, which offers a factor of $\sim$20 increase in area
compared to the WFC3 ERS programme {\color{black}(but typically in
fewer filters compared to the 10-band HST data used here)}. While
we have presented plausible evidence for the general
insignificance of major mergers in producing SGs and driving
stellar mass growth at high redshift, the role of such mergers in
the redshift range $1<z<3$ has to be quantified further, by
directly comparing the rate of emergence of SGs with the observed
major-merger rate. Again, this will benefit from better statistics
than is available in either the WFC3 ERS sample or the existing
CANDELS data. Bulk visual classification of galaxy morphologies,
required for the reliable identification of mergers and SGs (see
e.g. Kaviraj et al. 2007a; Kartaltepe et al. 2010) from large
datasets such as CANDELS, can be achieved using novel techniques
such as {\color{black}the} Galaxy Zoo project.

While rest-frame UV/optical photometry provides reasonable
constraints on galaxy star formation histories, they can be
(significantly) improved using forthcoming panchromatic data.
Spectroscopic line indices and/or radio {\color{black}continuum}
data can put strong priors on stellar/gas-phase metallicities and
the current star-formation rate respectively, enabling us to
significantly reduce uncertainties on the derived star formation
history parameters (c.f. Section 3). Intriguingly, the derived
star formation timescales in our SGs appear somewhat longer than
those implied by the high [$\alpha$/Fe] ratios observed in the
low-redshift Universe. However, photometric analyses such as the
one performed here typically yield large error bars on the derived
timescales, making it difficult to ascertain whether this is a
real effect. A systematic study of the stellar/gas-phase
metallicities, [$\alpha$/Fe] ratios and star formation rates of
massive SGs at $z>1$, using forthcoming
{\color{black}spectrographs} like KMOS and radio
{\color{black}continuum} surveys using the SKA precursors (e.g.
e-MERLIN), thus becomes a compelling exercise. Looking further
ahead, future morphological studies, using the JWST and the
extremely large telescopes, will enable us to probe massive
galaxies beyond $z>3$, bridging the gap between SGs at the epoch
of peak star formation and their progenitors.

{\color{black} In future papers, we will systematically tackle
these issues using the forthcoming datasets mentioned above, both
to further our understanding of the emerging SG population and, in
particular, to bring such empirical results to bear on our
emerging theoretical models for describing the high-redshift
Universe.}

\section*{Acknowledgements}
We are grateful to the referee Scott Trager for many constructive
comments that helped improve the original manuscript. Daniel
Thomas, Pieter van Dokkum, Ignacio Ferreras, Mariska Kriek,
Claudia Maraston, Simona Mei and Ewan Cameron are thanked for
comments and related discussions. SK is grateful for the generous
hospitality of the California Institute of Technology, where most
of this work was completed. SK also acknowledges fellowships from
Imperial College London, the Royal Commission for the Exhibition
of 1851 and Worcester College, Oxford.

This paper is based on Early Release Science observations made by
the WFC3 Scientific Oversight Committee. We are grateful to the
Director of the Space Telescope Science Institute for awarding
Director's Discretionary time and deeply indebted to the brave
astronauts of STS-125 for rejuvenating HST. {\color{black}Support
for HST program 11359 was provided by NASA through grant GO-11359
from the Space Telescope Science Institute, which is operated by
the Association of Universities for Research in Astronomy, Inc.,
under NASA contract NAS 5-26555. R.A.W. also acknowledges support
from NASA JWST Interdisciplinary Scientist grant NAG5-12460 from
GSFC. The work of AD has been partly supported by the ISF grant
6/08 by GIF grant G-1052-104.7/2009, DIP grant
STE1869/1-1.GE625/15-1 and NSF grant AST-1010033.}

\nocite{Tal2012,Newman2011,Ferreras2000,VD2011,Dekel2009b,Deharveng2000,Yi2005,Kaviraj2007a,Kaviraj2007b,Kaviraj2008,Kaviraj2011,delarosa2011,Kartaltepe2010,Rutkowski2012,Salim2010}


\bibliographystyle{mn2e}
\bibliography{references}


\end{document}